\documentclass{article}
\usepackage[T1]{fontenc}
\usepackage[utf8]{inputenc}
\usepackage{nonismir} 
\usepackage{amsmath,cite,url}
\usepackage{graphicx}
\usepackage{color,xcolor}

\usepackage{amssymb}
\usepackage{amsfonts}
\usepackage{multirow, booktabs, bbding, enumerate, enumitem}%
\usepackage{tabularx,array}
\usepackage{dblfloatfix}
\usepackage{cuted}
\usepackage{caption}

\newcolumntype{C}{>{\centering\arraybackslash}X}
\newcolumntype{R}{>{\raggedright\arraybackslash}X}
\newcolumntype{L}{>{\raggedleft\arraybackslash}X}

\newcommand{\ctoprule}[1]{%
  \noalign{\global\savedwidth=\cmidrulewidth
  \global\cmidrulewidth=\heavyrulewidth}%
  \cmidrule{#1}%
  \noalign{\global\cmidrulewidth=\savedwidth}%
}
\newlength{\savedwidth}

\title{Iterative Audio Separation with Mixture Consistency\\via MIMO Model Extension}

\multauthor
  {Yukara Ikemiya$^1$ \hspace{1cm} WeiHsiang Liao$^1$ \hspace{1cm} Yuki Mitsufuji$^1$}
  {$^1$ SonyAI\\
  {\tt\small \{yukara.ikemiya, weihsiang.liao, yuhki.mitsufuji\}@sony.com}
  }

\def\authorname{F. Author, S. Author, and T. Author}

\usepackage[bookmarks=false,pdfauthor={\authorname},pdfsubject={\pdfsubject},hidelinks]{hyperref}

\begin{document}

\maketitle

\begin{abstract}
This paper proposes a general framework for stable and effective iterative audio separation
with mixture consistency by extending source separation models to a multi-input multi-output (MIMO) configuration.
In the field of audio separation, mixture consistency is an essential property for many applications
that require accurate phase and timbral information of target sources.
While iterative approaches such as diffusion models
achieve perceptually superior results in speech enhancement or user-guided target source separation tasks,
most existing methods focus on single-step separation
with a single-input single-output (SISO) or single-input multi-output (SIMO) configuration
through architectural improvements,
since mixture-consistent audio separation is generally regarded as a regression problem that admits a unique solution.
By extending these architectures to a MIMO configuration,
we introduce iterative prediction without compromising the architectural advantages
or the characteristics of mixture consistency.
We conduct a comprehensive ablation study of combining the framework with discriminators
and extending it to a generative model.
Experimental results demonstrate significant performance improvements when applying the proposed framework
to state-of-the-art separation models.
The code and pretrained model weights are available at \url{https://github.com/SonyResearch/mimo-audio-separation}.
\end{abstract}

\section{Introduction}\label{sec:introduction}

Audio source separation is one of the fundamental tasks in music information retrieval (MIR)
and audio signal processing, with a wide range of applications including music production,
broadcasting, and audio dataset construction.
Given a mixture signal, the objective is to decompose it into its constituent source signals,
an inherently ill-posed problem that has been studied extensively.

Recently, deep learning has driven remarkable progress in this field.
Generative frameworks including diffusion-based methods have achieved
perceptually superior results in tasks such as 
speech enhancement/separation~\cite{DiffSE2021,SGMSE2023,DiffSep2023,FlowSE2025,FLOSS2025,Geneses2026},
user-guided target source extraction~\cite{FlowSep2025,Wen2025},
and generative music separation~\cite{MSDM2024,SMGLDM2025}.
By leveraging their ability to model complex prior distributions over target sources,
these models produce natural-sounding outputs, even under challenging acoustic conditions.
The iterative inference mechanism is one of the key contributors to their success,
allowing the model to progressively refine its estimates over multiple denoising steps.

Despite this progress, audio separation with \textit{mixture consistency}, where the
separated sources are constrained to sum back to the original mixture, remains essential 
for many practical applications such as stem-based remixing, music restoration,
and format conversion.
In these tasks, large deviations from the original mixture are undesirable
and the phase and timbral characteristics of each source must be precisely preserved
to avoid audible artifacts in downstream processing.
Beyond these conventional applications,
it has also become an important tool for constructing training data for generative models.
The proliferation of powerful music generation and editing systems has created
a growing demand for high-quality multi-stem datasets.
Since large-scale multi-track recordings with ground-truth stems are scarce,
source separation is increasingly employed to derive pseudo-stem data from
existing mixed recordings~\cite{SingSong2023,FastSAG2024,MusicGenStem2025,AnyAccomp2026},
enabling the training of controllable and steerable generative systems at scale.

\begin{figure}[tb]
  \centering
  \includegraphics[alt={MIMO model extension},width=0.99\linewidth]{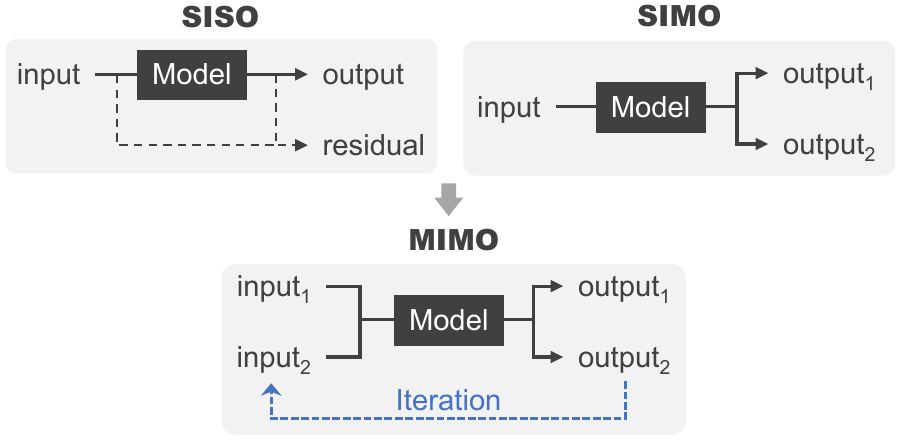}
  \caption{MIMO model extension.}
  \label{fig:mimo_ext}
  \vspace{-0.5em}
\end{figure}

\begin{figure*}[tb]
  \centering
  \includegraphics[alt={Overview of the proposed framework for iterative audio separation},width=0.98\linewidth]{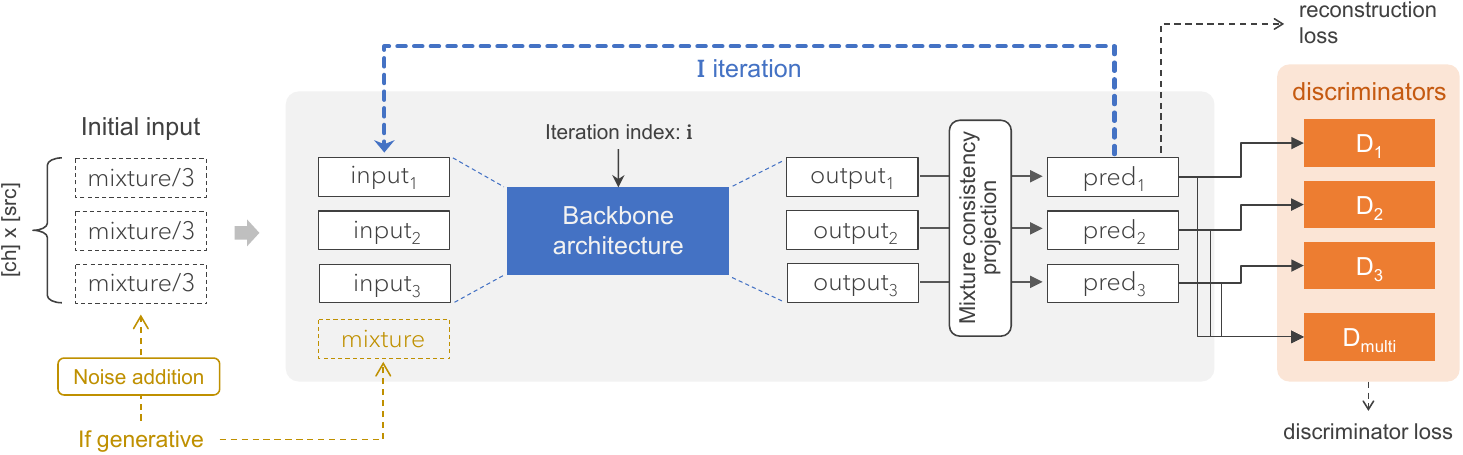}
  \caption{Overview of the proposed framework for iterative audio separation. (A 3-source case)}
  \label{fig:overview}
  \vspace{-0.6em}
\end{figure*}

\begin{figure}[tb]
  \centering
  \includegraphics[alt={Mask weighted prediction for spectral masking-based methods},width=0.88\linewidth]{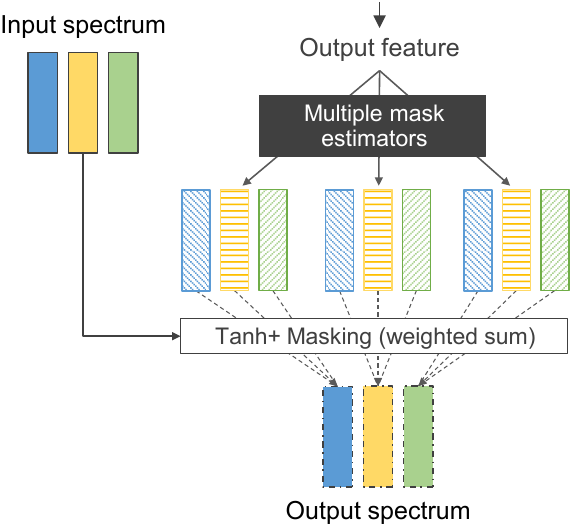}
  \caption{Mask weighted prediction for spectral masking-based methods.}
  \label{fig:mask_pred}
  \vspace{-0.5em}
\end{figure}

Since mixture-consistent audio separation is regarded as a regression problem that admits a unique solution,
most existing methods focus on single-step inference using
single-input single-output (SISO) or single-input multi-output (SIMO) architectures,
where a mixture is processed in a single forward pass to predict
target sources~\cite{ConvTasNet2019,HDemucs2021,BSRNN2023,BSRoformer2023,Melroformer2023,SCNet2024}.
These methods have benefited greatly from advances in neural network design,
yet they leave the power of \textit{iterative refinement} largely untapped.
However, incorporating iterative refinement into mixture-consistent separation is non-trivial.
Diffusion-based models pursue a distributional objective through iterative sampling,
which is fundamentally incompatible with enforcing strict mixture consistency at every step.
Naively running multi-round inferences with existing SISO/SIMO models also fails
because these models cannot correct their own errors in their previous outputs,
leading to error accumulation and over-separation rather than genuine refinement.

To bridge this gap, this paper proposes a general framework for stable and effective
iterative audio separation with mixture consistency by extending separation
models to a multi-input multi-output (MIMO) configuration (Fig.~\ref{fig:mimo_ext}).
The key insight is that by feeding back all the current source estimates as inputs
of the next step, the model can mutually refine all sources while
preserving mixture consistency at each step.
This extension requires only minor modifications to existing architectures,
making it broadly applicable to arbitrary separation models.
We further explore how this framework interacts with adversarial training
and how it can be extended toward a generative formulation.
Experimental results on music source separation tasks demonstrate consistent and
significant improvements over strong single-step baselines.

\section{Related work}
\label{sec:related_works}

Iterative source separation has been studied most explicitly
in the context of multi-speaker speech separation.
Wang et al. \cite{UnfoldedMISI2018} incorporate iterative phase reconstruction into
an end-to-end trainable architecture and impose mixture
reconstruction constraints during the iterative process.
More recently, Scheibler et al. \cite{FLOSS2025} formulate speech separation as a constrained
flow-matching problem and enforces mixture consistency throughout
the generative process.

Another line of work explores iterative refinement without explicitly enforcing mixture consistency.
Koyama et al. \cite{DEQ-UMX2022} introduce deep equilibrium models into music source separation
and show that implicit fixed-point-style computation is effective for separation.
Bai et al. \cite{TSBSMAMBA2024} adopt an explicit two-stage sequential architecture
in which the second stage predicts residual mappings to recover details missed by the first-stage estimate.
Zang et al. \cite{Zang2025} show that a pretrained one-step separator can be iteratively reused
at inference time without additional training, leading to improved separation quality. 
While these methods support the general view that
source separation can benefit from iterative inference,
they do not jointly refine all sources at each step,
which can make the refinement process more susceptible to over-separation
and limit the gains obtainable from further iteration.

\section{Proposed framework}
\label{sec:proposed_framework}

We propose a general framework for iterative audio separation with mixture consistency
based on a multi-input multi-output (MIMO) extension of separation architectures (Fig.~\ref{fig:overview}).
The core idea is to feed back the current source estimates as inputs to the model
at each iteration, enabling the model to progressively refine its predictions
while preserving mixture consistency throughout.

\subsection{MIMO Architecture for Iterative Separation}
\label{subsec:mimo}

Let $\mathbf{x} \in \mathbb{R}^{C \times T}$ denote the mixture signal with $C$ channels
and $T$ time samples, and let $\mathbf{s}_k \in \mathbb{R}^{C \times T}$
($k = 1, \ldots, K$) denote the $k$-th source, satisfying
$\mathbf{x} = \sum_{k=1}^{K} \mathbf{s}_k$.
Conventional models take the mixture as input and estimate sources
in a single forward pass:
$(\hat{\mathbf{s}}_1, \ldots, \hat{\mathbf{s}}_K) = f(\mathbf{x})$.

In the proposed MIMO framework, the model receives all the source estimates
from the previous iteration as inputs.
At iteration $i$, the update rule becomes
\begin{equation}
  (\hat{\mathbf{s}}_1^{(i)}, \ldots, \hat{\mathbf{s}}_K^{(i)})
  = f\!\left(i;\, \hat{\mathbf{s}}_1^{(i-1)};\, \ldots;\, \hat{\mathbf{s}}_K^{(i-1)}\right),
  \label{eq:mimo_update}
\end{equation}
where the time step $i$ is explicitly provided as a condition to the model.
Since the input to the first iteration must also satisfy mixture consistency,
the source inputs are initialized as $\hat{\mathbf{s}}_k^{(0)} = \mathbf{x} / K$ for all $k$.

Because the extension amounts only to widening the input and output channel dimensions from $C$
to $K \cdot C$, virtually any SISO or SIMO architecture can be converted to MIMO
without significant structural modification beyond adapting the input and output modules,
which makes the proposed framework broadly applicable to state-of-the-art models.
While some models directly output the estimated signals~\cite{HDemucs2021,SCNet2024},
another common type of model works on the complex spectrogram, predicts a complex mask for each source,
and applies it to the mixture spectrogram~\cite{BSRNN2023,BSRoformer2023,Melroformer2023}.
For such mask-based models, the MIMO extension is realized by predicting
a separate mask for each combination of an input and an output estimate,
and computing each output estimate as a weighted sum of all inputs using the predicted masks (Fig.~\ref{fig:mask_pred}).
Specifically, the $k$-th output source spectrum at iteration $i$ is given by
\begin{equation}
  \hat{\mathbf{S}}_k^{(i)} = \sum_{j=1}^{K} \mathbf{m}_{k,j}^{(i)} \odot \hat{\mathbf{S}}_j^{(i-1)},
  \label{eq:masking_mimo}
\end{equation}
where $\mathbf{m}_{k,j}^{(i)}$ is the mask for the $k$-th output source and 
the $j$-th input source, and $\hat{\mathbf{S}}_j^{(i-1)}$ is
the spectral representation of the $j$-th input source from the previous iteration.

\subsection{Mixture Consistency Projection}
Mixture consistency is enforced after each iteration by mixture consistency projection~\cite{MCP2019,USS2019},
\begin{equation}
  \hat{\mathbf{s}}_k^{(i)} \leftarrow \hat{\mathbf{s}}_k^{(i)}
  + \frac{1}{K}\!\left(\mathbf{x} - \sum_{j=1}^{K} \hat{\mathbf{s}}_j^{(i)}\right).
  \label{eq:mc_proj}
\end{equation}
Beyond enforcing the constraint strictly,
this projection also restricts the search space of possible signals
the model must predict at each step,
thereby stabilizing training and improving separation performance.

\subsection{Discriminators}
\label{subsec:disc}

Discriminators can optionally be incorporated into the framework
to further enhance the separation performance.
We investigate two complementary types.

\noindent\textbf{Stem-wise discriminator:}
A stem-wise discriminator is trained to distinguish each estimated signal $\hat{\mathbf{s}}_k^{(i)}$
from the corresponding ground-truth signal $\mathbf{s}_k$.
It captures the nuances of individual stems and has been reported
to contribute to improved separation performance~\cite{Disc2018,Disc2022}.

\noindent\textbf{Multi-stem discriminator:}
A multi-stem discriminator is trained to distinguish a set of
estimated signals $\{\hat{\mathbf{s}}_k^{(i)}\}$
from a set of ground-truth signals $\{\mathbf{s}_k\}$.
This is functionally equivalent to the context-based discriminator in~\cite{Disc2023}.

While a feature-matching loss is commonly used in adversarial training for generative tasks,
we found empirically that incorporating this loss causes training instability in our framework.
Therefore, only the adversarial loss from each discriminator is used in this paper.

\subsection{Generative Extension}
\label{subsec:gen}

While the core MIMO framework is inherently discriminative, it can be extended
to incorporate generative characteristics by introducing stochastic refinement
through noise perturbation during training and inference.
The motivation is to enable the model to explore a distribution of plausible separations
rather than converging to a single deterministic solution,
which can help escape local optima and improve robustness to challenging acoustic conditions.

Specifically, we perturb the input mixture with random noise
while preserving mixture consistency through the \textit{zero-sum noise} scheme proposed in~\cite{FLOSS2025}.
This constructs noise vectors $\boldsymbol{\epsilon}_k$ satisfying $\sum_{k=1}^{K} \boldsymbol{\epsilon}_k = \mathbf{0}$,
ensuring that the perturbed input remains mixture-consistent.
The model is then trained to denoise the perturbed mixture back to clean sources,
which is analogous to the reverse process in diffusion models.

To further leverage prior information from the clean mixture $\mathbf{x}$,
we supply it as an additional input at every iteration, as depicted in Fig.~\ref{fig:overview}.
This extension enables the iterative refinement process to combine
the generative benefits with the discriminative constraint of mixture consistency.

\subsection{Input Audio Normalization}
\label{subsec:norm}

To improve training stability and overall performance,
the input mixture is peak-normalized before being fed into the model.
Specifically, given the mixture $\mathbf{x}$,
the input to the model ($\tilde{\mathbf{x}}$) is computed as
\begin{equation}
  \tilde{\mathbf{x}} = \mathbf{x} / \mathrm{scale}, \quad \mathrm{scale} = \max_t |\mathbf{x}[t]|.
  \label{eq:norm}
\end{equation}
At inference time, the output estimates are rescaled back to the original amplitude
by multiplying by $\mathrm{scale}$.


\begin{table*}[t!] 
\centering
\begin{tabularx}{\textwidth}{l|c|CCCCC|CCCC}
\toprule
    & & \multicolumn{5}{c}{Vocal} & \multicolumn{4}{c}{Accompaniment} \\
    Variant &  Params (M) & SDR & SIR & SAR & LMD & LPR & SDR & SIR & SAR & LMD\\
\midrule
    (1) SISO (BS-RoFormer) & 12.1 & 10.38 & 19.12 & 11.74 & 1.57 & -44.7 & 21.92 & 35.51 & 22.94 & 0.39\\
    (2) SIMO & 12.1 & 10.41 & 19.91 & 11.53 & 1.48 & -41.6 & 20.98 & 34.72 & 22.02 & 0.39\\
    (3) MIMO (I=1) & 16.1 & 10.36 & 19.70 & 11.70 & 1.57 & -45.2 & 20.61 & 33.45 & 21.70 & 0.39\\
    (4) 3 + disc[stem] & 16.1 (1.0) & 10.45 & 19.68 & 11.77 & 1.31 & -45.3 & 21.74 & 34.78 & 22.78 & \textbf{0.32}\\
    (5) 3 + disc[multi] & 16.1 (0.5) & 10.37 & 19.76 & 11.59 & 1.53 & -44.9 & 21.64 & 35.09 & 22.78 & 0.37\\
    (6) 3 + disc[stem,multi] & 16.1 (1.5) & 10.36 & 19.64 & 11.63 & 1.40 & -44.2 & 21.63 & 35.39 & 23.16 & 0.33\\
\midrule   
    (7) MIMO (I=3) & 16.3 & 10.72 & 20.84 & 11.89 & 1.59 & \underline{-45.4} & 22.07 & 35.97 & 22.90 & 0.39\\
    (8) 7 + disc[stem] & 16.3 (1.0) & \textbf{11.05} & \underline{21.29} & \underline{12.14} & \textbf{1.23} & -43.4 & \underline{22.86} & \underline{37.08} & \underline{23.77} & \textbf{0.32}\\
    (9) 7 + disc[stem,multi] & 16.3 (1.5) & 10.62 & 20.57 & 11.79 & 1.31 & -41.4 & 21.93 & 35.89 & 23.01 & 0.34\\
    (10) 8 + generative & 20.2 (1.0) & \underline{11.00} & \textbf{21.35} & \textbf{12.16} & \textbf{1.23} & \textbf{-47.5} & \textbf{23.21} & \textbf{37.47} & \textbf{24.26} & \textbf{0.32}\\
\bottomrule
\end{tabularx}
\caption{Evaluation results of ablation study.\
Higher SDR/SIR/SAR and lower LMD/LPR represent better performance.\
The parameter counts in ($\cdot$) denote those of discriminators,\
while the main parameter counts exclude discriminators.
Bold and underline indicate the \textbf{best} and \underline{second-best} values, respectively.
}
\label{tab:ablation}
\end{table*}

\begin{figure*}[tb]
  \centering
  \includegraphics[alt={Overview},width=0.99\linewidth]{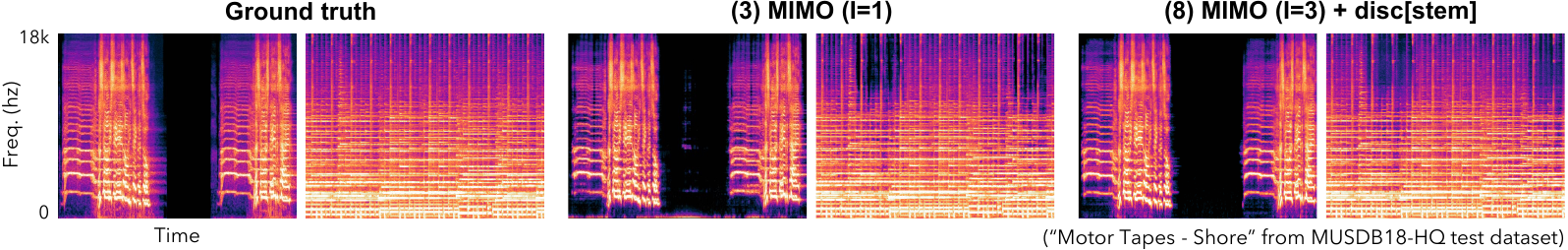}
  \caption{Samples of vocal-accompaniment separation from the ablation study.}
  \label{fig:separation_samples}
  \vspace{-1.0em}
\end{figure*}

\subsection{Loss Function}
\label{subsec:loss}

The loss at iteration $i$ combines a reconstruction term and an adversarial term:
\begin{equation}
  \mathcal{L}_i = \mathcal{L}_i^{\text{rec}}
  + \lambda_{\text{adv}}\, \mathcal{L}_i^{\text{adv}},
  \label{eq:loss_iter}
\end{equation}
where $\mathcal{L}_i^{\text{adv}}$ is the average loss over all discriminators.
The reconstruction loss combines L1 loss in both the time domain and the complex spectral domain,
following~\cite{BSRNN2023,BSRoformer2023}.
\begin{equation}
  \mathcal{L}_i^{\text{rec}} = \frac{1}{K}\sum_{k=1}^{K} \left(
  \|\hat{\mathbf{S}}_k^{(i)} - \mathbf{S}_k\|_1
  + \lambda_{\text{wav}} \|\hat{\mathbf{s}}_k^{(i)} - \mathbf{s}_k\|_1
  \right),
    \label{eq:loss_recon}
\end{equation}
where $\mathbf{S}_k$ denotes the short-time Fourier transform (STFT) representation of $\mathbf{s}_k$.
To encourage monotonic refinement across iterations, the final training objective
accumulates losses over all $I$ iterations with per-iteration weights $\lambda_i$:
\begin{equation}
  \mathcal{L} = \sum_{i=1}^{I} \lambda_i\, \mathcal{L}_i.
  \label{eq:loss_total}
\end{equation}

\subsection{Relationship to Fixed-Point Iteration}
\label{subsec:fpi}

The iterative update in Eq.~\eqref{eq:mimo_update} can be interpreted as
approximately solving the fixed-point equation
$f\!\left(\{\mathbf{s}_k\}_{k=1}^K\right) = \{\mathbf{s}_k\}_{k=1}^K$,
where the model is trained to converge to a self-consistent solution
that satisfies mixture consistency.
Koizumi et al.~\cite{Wavefit2022} explored a related idea in the context of
neural vocoding, where fixed-point-style iterative training enables
a lightweight model to achieve high-quality waveform synthesis in only a small number
of refinement steps.
The proposed framework brings this principle to mixture-consistent source separation,
combining the inductive bias of fixed-point training with the architectural strengths
of modern separation models.

\section{Experiments}
\label{sec:experiments}

\subsection{Dataset}

We use the MUSDB18-HQ~\cite{MUSDB18HQ} dataset, which consists of 150 professionally produced stereo music tracks
with four annotated sources: \textit{vocals}, \textit{bass}, \textit{drums}, and \textit{other}.
The dataset is split into 100 tracks for training and 50 tracks for testing.
The audio is provided in uncompressed format at a sampling rate of 44.1 kHz.
For vocal-accompaniment separation, the accompaniment sound is constructed
by summing the \textit{bass}, \textit{drums}, and \textit{other} sources.

\subsubsection{Data augmentation}

For each stem in every training sample, we extract an audio segment from a random position in a random track
while avoiding complete silence.
Additionally, we apply channel shuffling, phase inversion, and random gain variations as augmentations.
With a small probability, each stem is replaced with a silent signal.

\subsubsection{Test data preprocessing}

The MUSDB18-HQ test data contains undesired artifacts such as small DC components,
floor noise, and leakage from other stems during silent intervals.
Since these artifacts are not relevant to source separation,
including them in evaluation hinders an accurate assessment of separation performance.
To enable more reliable evaluation, we apply the test data preprocessing proposed in \cite{Tunturi2025}.
Specifically, each stem is divided into 1-second segments,
and regions where the maximum amplitude falls below 0.01 are replaced with silence.
This approach effectively removes most artifacts and also enables evaluation during silent intervals,
which are often ignored in previous work.

%
%

\begin{table*}[!htbp] 
\addtocounter{table}{1}
    \centering
    \resizebox{0.85\textwidth}{!}{
    \begin{tabular}[t]{l|c|ccccc|cccc}
        \multicolumn{2}{c}{} & \multicolumn{9}{c}{MUSDB18-HQ (test)} \\
    \toprule
        & & \multicolumn{5}{c|}{Vocal} & \multicolumn{4}{c}{Accompaniment} \\
        Model &  Params (M) & SDR & SIR & SAR & LMD & LPR & SDR & SIR & SAR & LMD\\
    \midrule
        BS-R(rep) & 72.2 & 11.92 & 21.75 & 13.07 & 1.51 & -48.2 & 24.54 & \textbf{40.15} & 25.47 & 0.37 \\
        \textbf{MIMO BS-R} & 71.4 (1.0) & \textbf{12.29} & \textbf{23.93} & \textbf{13.71} & \textbf{1.07} & \textbf{-58.4} & \textbf{26.01} & 39.85 & \textbf{26.73} & \textbf{0.30} \\
    \bottomrule
        \multicolumn{2}{c}{} & \multicolumn{9}{c}{\raisebox{0pt}[3ex][0pt]{}MoisesDB (OOD)} \\
    \ctoprule{3-11}
        \multicolumn{2}{c}{} & 12.15 & 21.89 & 12.96 & 1.27 & -54.4 & 21.27 & 32.58 & 22.08 & 0.43 \\
        \multicolumn{2}{c}{} & \textbf{12.44} & \textbf{23.82} & \textbf{13.09} & \textbf{1.10} & \textbf{-55.7} & \textbf{21.40} & \textbf{33.97} & \textbf{22.09} & \textbf{0.34} \\
    \ctoprule{3-11}
    \end{tabular}
    }
    \captionof{table}{Detailed comparison between BS-RoFormer and MIMO BS-RoFormer.}
    \label{tab:bsr_comparison}
\vspace{-0.7em}
\end{table*}

\begin{table}[t!]
\addtocounter{table}{-2}
    \centering
    \resizebox{1.0\linewidth}{!}{
    \begin{tabular}{cccccc}
    \toprule
        \multicolumn{2}{c|}{BS-R} & \multicolumn{2}{c|}{Mel-R} & \multicolumn{1}{c|}{BS-R(rep)} & \textbf{MIMO BS-R} \\
    \midrule
        72.2M & 82.8M & 84.2M & 94.8M & 72.2M & 71.4 (1.0)M \\
    \midrule
        10.78 & 11.02 & 11.21 & 11.60 & 11.40 & 11.62 \\
    \bottomrule
    \end{tabular}
    }
    \caption{Vocal SDR evaluation via \texttt{museval}.}
    \label{tab:sdr_museval}
\vspace{-1.2em}
\end{table}

\addtocounter{table}{1} 

\subsection{Model Details}

We adopt \textbf{BS-RoFormer}~\cite{BSRoformer2023},
which is one of the state-of-the-art models for music source separation,
as a backbone model for our ablation study (Sec.~\ref{sec:ablation})
and vocal-accompaniment separation (Sec.~\ref{sec:va_sep}).
For the MIMO extension, we increase the number of input channels in the band-split module
by a factor of the number of stems, and implement multi-output via the multiple mask prediction 
scheme described in Sec.~\ref{subsec:mimo}.
The time step condition is provided as a learnable embedding prepended to the input features.
For the 4-stem separation experiment (Sec.~\ref{sec:4stem_sep}), we adopt \textbf{SCNet}~\cite{SCNet2024},
which achieves state-of-the-art performance with a small parameter count for this task.
Since SCNet is originally designed in SIMO format, 
it can be extended to MIMO by multiplying the input channel dimension by the number of stems.
BS-RoFormer and SCNet variants are trained for 1 million steps with batch sizes of 96 and 48
using bf16 and fp32 precision, respectively.

Through empirical observation, we found that performance improvements saturate around 4 iterations.
Considering the computational cost, we set the number of iterations $I$ to 3 and 2
for MIMO BS-RoFormer and MIMO SCNet, respectively.
Hyperparameters are configured as follows: $\lambda_{\text{adv}} = 0.01$ and $\lambda_{\text{wav}} = 2.0$.
The per-iteration loss weights $\lambda_i$ are set to $i$ to emphasize performance
in later iterations for monotonic refinement.

For discriminators, we adopt the multi-band discriminator design from~\cite{DAC2023},
which has proven effective in audio compression tasks for capturing features across a wide frequency range.
Both stem-wise and multi-stem discriminators employ this design with an FFT size of 2048,
hop size of 512, and frequency bands divided into 5 sub-bands.

For inference, we employ overlap-add processing
with a hop size of half the input length.
Overlapping regions are blended with linear crossfade to ensure smooth transitions.

\subsection{Evaluation Metrics}

Source-to-distortion ratio (\textbf{SDR}), source-to-interference ratio (\textbf{SIR}), 
and source-to-artifact ratio (\textbf{SAR})~\cite{SDR2006} are used for evaluation.
Additionally, we compute the log-magnitude distance (\textbf{LMD}) in the STFT domain~\cite{Disc2022} to evaluate timbral fidelity.
These metrics are computed only on non-silent intervals for each stem.

To evaluate leakage and artifacts in silent intervals,
we introduce the metric proposed in~\cite{Tunturi2025} as the leakage power ratio (\textbf{LPR}).
For each stem, the median energy of the ground-truth non-silent frames is first computed as a reference level.
Then, for the estimated signal, the energy of the frames corresponding to silent segments in the ground-truth signal is computed
and divided by the reference level.
The value is converted to decibels to yield the LPR metric.
A lower LPR indicates less leakage and artifacts in the silent intervals of the estimated signal.
For the "accompaniment" stem, which contains very few or no silent intervals,
the LPR metric is not computed.

For each track, metrics are computed by dividing the audio into 1-second segments
and taking the median of the metrics across all segments.
The final metrics are obtained by averaging the metrics across all tracks.

\begin{figure}
  \centering
  \includegraphics[alt={SDR over iterations},width=0.99\linewidth]{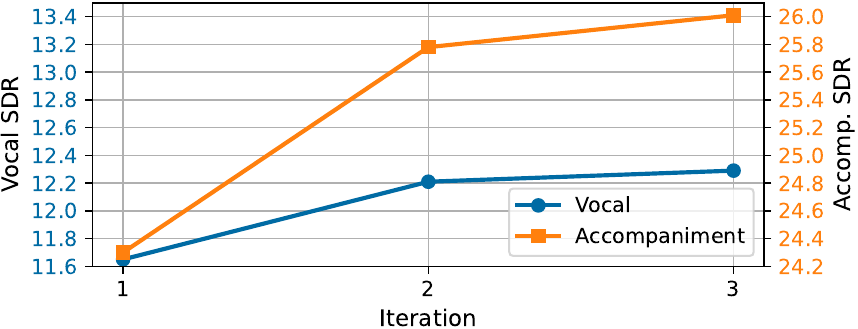}
  \caption{SDR over iterations.}
  \label{fig:sdr_iter}
\vspace{-1.2em}
\end{figure}

\begin{table*}[!htbp] 
\centering
\resizebox{1.0\linewidth}{!}{
\begin{tabular}{l|c|ccc|ccc|ccc|ccc}
\toprule
    & & \multicolumn{3}{c|}{Vocal} & \multicolumn{3}{c|}{Bass} & \multicolumn{3}{c|}{Drums} & \multicolumn{3}{c}{Other} \\
    Model &  Params (M) & SDR & SIR & SAR & SDR & SIR & SAR & SDR & SIR & SAR & SDR & SIR & SAR\\
\midrule
    SCNet & 10.6 & 10.39 & 18.15 & 11.77 & 9.10 & 15.64 & 12.47 & 11.32 & 19.58 & 12.72 & 7.21 & 12.81 & 9.58 \\
    \textbf{MIMO SCNet} & 10.6 (0.5) & 10.57 & 18.62 & 11.92 & \textbf{9.80} & \textbf{16.98} & \textbf{13.65} & \textbf{11.73} & \textbf{20.39} & \textbf{12.94} & \textbf{7.63} & \textbf{13.61} & \textbf{9.75} \\
    \textbf{MIMO SCNet-gen} & 10.6 (0.5) & \textbf{10.72} & \textbf{18.75} & \textbf{12.05} & 9.73 & 16.86 & 13.29 & 11.61 & 20.08 & 12.92 & 7.61 & 13.46 & 9.67 \\
\bottomrule
\end{tabular}
}
\captionof{table}{Evaluation results of 4-stem separation.}
\label{tab:4stem_comp}
\vspace{-1.0em}
\end{table*}

\subsection{Ablation Study}
\label{sec:ablation}

To validate the effectiveness of the proposed framework and its components
including iterative prediction, discriminators, and generative extension,
we first conduct a comprehensive ablation study on vocal-accompaniment separation.
We start with a basic comparison of SISO, SIMO, and MIMO (I=1) to isolate the effects
of the input-output format.
We then evaluate discriminators in isolation using MIMO (I=1),
followed by a final comparison combining discriminators and generative extension with MIMO (I=3).
The SISO variant corresponds to the original BS-RoFormer baseline.
The additional parameters in MIMO variants arise solely from the increased number of mask predictors;
the core Transformer architecture remains identical across all variants
(Transformer dimension/heads: 192/8, blocks: 4, time/frequency module depth: 1)
and all variants are trained with 4-second audio input.

Table~\ref{tab:ablation} shows the overall results of the ablation study.
Regarding 1-step prediction, SISO, SIMO, and MIMO tend to yield similar performance across most metrics.
This suggests that, even in the SISO setting, separating a 2-source mixture implicitly
estimates the residual audio alongside the target,
so the differences between output configurations have limited impact.
Regarding the effect of discriminators,
the stem-wise discriminator consistently improves performance
whereas adding the multi-stem discriminator tends to degrade performance.
This pattern holds consistently for both 1- and 3-step cases,
suggesting that the multi-stem discriminator conflicts with the separation objective,
hindering rather than helping the model.

The largest gains come from iterative prediction.
Iterative prediction yields consistent improvements across all metrics,
and further combining it with the stem-wise discriminator delivers better performance.
The generative variant reduces artifacts in the accompaniment source,
showing additional benefits for overall separation performance.
Fig.~\ref{fig:separation_samples} shows separation samples of the single-step MIMO variant (3)
and the iterative MIMO variant (8).
Iterative prediction with the discriminator effectively reduces accompaniment leakage in the vocal source
while mitigating the loss of high-frequency components in the accompaniment.

\subsection{Vocal-Accompaniment Separation}
\label{sec:va_sep}

We conduct comparative experiments on larger-scale models for vocal-accompaniment separation.
While the generative variant shows superior performance,
we use the predictive variant (8) here
to more directly evaluate the performance of the iterative framework.
As baseline comparisons, we reference results for BS-RoFormer ('\textbf{BS-R}')
and Mel-RoFormer ('\textbf{Mel-R}') from \cite{Melroformer2023}.
To ensure fair comparison, we also train BS-RoFormer from scratch
using the same training pipeline and input audio normalization employed in this work,
denoted as '\textbf{BS-R(rep)}'.
The proposed framework ('\textbf{MIMO BS-R}') adopts the same Transformer architecture
as the 72.2M BS-RoFormer baseline,
but adjusts the hidden feature dimensions in the mask predictors
from the original scale of 4 to 1.5 to maintain a comparable parameter size,
accounting for the increased number of mask predictors.
These variants are trained with 8-second audio input, consistent with the original papers.

Table~\ref{tab:sdr_museval} shows the SDR comparison with published results.
For this comparison, we compute SDR using \texttt{museval}~\cite{MUSEVAL2018}
without test data preprocessing,
consistent with the original papers, with final scores computed as the median across all tracks.
The results demonstrate substantial improvements from the proposed MIMO framework.
MIMO BS-R achieves 11.62 dB, a 0.22 dB improvement over BS-R(rep).
This gain is notable given that the baseline BS-R(rep) already shows improved performance
relative to the published BS-R result (10.78 dB),
suggesting that our training procedure—including data augmentation and input audio normalization—
contributes to the overall performance.
When compared against published results for larger models,
MIMO BS-R achieves competitive performance with fewer parameters,
underscoring the efficiency of the proposed approach.

Table~\ref{tab:bsr_comparison} provides a detailed comparison
between BS-RoFormer and MIMO BS-RoFormer across all evaluation metrics.
We additionally evaluate on MoisesDB~\cite{Moisesdb2023} comprising 240 multi-stem tracks
as an out-of-domain (OOD) dataset to assess generalization performance.
The results show that MIMO BS-RoFormer consistently outperforms
the BS-RoFormer baseline across all metrics for both vocal and accompaniment.
On the OOD dataset, MIMO BS-RoFormer demonstrates substantial improvements in vocal SDR, SIR, and LMD, 
while maintaining comparable performance on other metrics.
Fig.~\ref{fig:sdr_iter} shows the SDR over iterations of MIMO BS-RoFormer and
demonstrates consistent gains across all iterations for both stems,
suggesting that the sources are progressively refined
through mutual interaction at each iteration.

\subsection{4-stem Separation}
\label{sec:4stem_sep}

We conduct a 4-stem separation experiment (vocals, bass, drums, and other)
using SCNet~\cite{SCNet2024} as a backbone model.
Following the original paper, we replace the reconstruction loss in Eq.~\eqref{eq:loss_recon}
with RMSE loss on the complex-valued spectrogram.
To improve training efficiency, we reduce the size of each stem-wise discriminator to 0.12M parameters.
For fair comparison, the SCNet baseline is also trained from scratch using the same training pipeline as MIMO SCNet.
These variants are trained with 11-second audio input, consistent with the original paper.

Table~\ref{tab:4stem_comp} reports the SDR, SIR, and SAR metrics for each stem.
The results show that MIMO SCNet consistently outperforms the SCNet baseline across all stems,
demonstrating the effectiveness of mutual refinement under mixture consistency
for a more complex separation case.
We additionally evaluate the generative variant ('\textbf{MIMO SCNet-gen}'),
which is trained with the same recipe as the 'MIMO SCNet' variant.
MIMO SCNet-gen achieves the best SDR, SIR, and SAR on the vocal stem,
while performing comparably to MIMO SCNet on the remaining stems,
indicating that the generative extension is beneficial for certain stems.

\section{Conclusion}

This paper proposes a general framework for iterative audio separation with mixture consistency
by extending source separation models to a multi-input multi-output (MIMO) configuration.
Experimental results show consistent and significant improvements
across multiple metrics compared to state-of-the-art single-step baselines.
The framework's broad applicability is demonstrated through experiments
with different backbone architectures and separation tasks.
Future work includes extending the framework to other separation tasks such as speech enhancement
and training on larger datasets to further improve performance and generalization.

\clearpage

\bibliography{2026_mimo-separation}

\end{document}